\documentstyle[12pt,epsfig,epstopdf,subfigure,color,amsfonts,amssymb,inputenc,graphicx,amsmath,amsxtra,amstext,slashed]{article}

\newcommand{\beq}{\begin{equation}}
\newcommand{\eeq}{\end{equation}}
\newcommand{\bea}{\begin{eqnarray}}
\newcommand{\eea}{\end{eqnarray}}
\newcommand{\al}{\alpha}
\newcommand{\be}{\beta}

\newcommand{\der}{\partial}

\newcommand{\m}{\mu}
\newcommand{\n}{\nu}

\newcommand{\nn}{\nonumber}
\newcommand{\ba}{\begin{eqnarray}}
\newcommand{\ea}{\end{eqnarray}}

\begin{document}

\baselineskip=15.5pt \pagestyle{plain} \setcounter{page}{1}
%
%--------+---------+---------+---------+---------+---------+---------+
%--------+---------+---------+---------+---------+---------+---------+
\begin{titlepage}

\vskip 0.8cm

\begin{center}

{\Large \bf  $1/N$ corrections to $F_1$ and $F_2$ structure
functions of vector mesons from holography}

\vskip 1.cm

{\large {{\bf Nicolas Kovensky$^{a,b,}$}{\footnote{\tt
nico.koven@fisica.unlp.edu.ar}}, {\bf Gustavo
Michalski$^{a,}$}{\footnote{\tt michalski@fisica.unlp.edu.ar}}, {\bf
and \\
Martin Schvellinger$^{a,}$}{\footnote{\tt
martin@fisica.unlp.edu.ar}}}}

\vskip 1.cm

{\it $^a$ Instituto de F\'{\i}sica La Plata-UNLP-CONICET. \\
Boulevard 113 e 63 y 64, (1900) La Plata, Buenos Aires, Argentina \\
and \\
Departamento de F\'{\i}sica, Facultad de Ciencias Exactas,
Universidad Nacional de La Plata. \\
Calle 49 y 115, C.C. 67, (1900) La Plata, Buenos Aires, Argentina.} \\

\vskip 1.cm

{\it $^b$ Institut de Physique Th\`{e}orique, CEA/Saclay 91191
Gif-sur-Yvette Cedex, France.}

\vspace{1.cm}

{\bf Abstract}

\vspace{1.cm}

\end{center}

The structure functions $F_1$ and $F_2$ of the hadronic tensor of
vector mesons are obtained at order $1/N$ and strong coupling using
the gauge/gravity duality. We find that the large $N$ limit and the
high energy one do not commute. Thus, by considering the high energy
limit first, our results of the first moments of $F_1$ for the rho
meson agree well with those from lattice QCD, with an important
improvement of the accuracy with respect to the holographic dual
calculation in the planar limit.

\noindent

\end{titlepage}

\newpage

{\small \tableofcontents}

\newpage

%%%%%%%%%%%%%%%%%%%%%%%%%%%%%%%%%%%%%%%%%%%%%%%%%%%%%%%%%%%%%%%%%%%%%%%%

%---------------------------------------------------------------------
%
\section{Introduction}
%
%---------------------------------------------------------------------

The idea of the present work is to investigate the leading $1/N$
corrections to the structure functions $F_1$ and $F_2$ of the
hadronic tensor of unpolarized vector mesons at strong 't Hooft
coupling $\lambda$, using the gauge/gravity duality. For this
purpose we consider vector mesons from the D3D7-brane system in type
IIB string theory \cite{Kruczenski:2003be}.

We are interested in the electromagnetic deep inelastic scattering
(DIS) of a charged lepton from a vector meson. The DIS cross section
is given by the contraction of a leptonic tensor, $l^{\mu \nu}$,
with a hadronic one, $W_{\mu \nu}$. The process involves an incoming
charged lepton interacting with a hadron with momentum $P$ through
the exchange of a virtual photon with momentum $q$, with the
condition $q^2 >> -P^2$. We consider the definitions given in
\cite{Hoodbhoy:1988am}, however we use the mostly-plus signature.
Thus, the DIS differential cross section is given by
\ba
\frac{d^2\sigma}{dx \ dy \ d\phi}&=&\frac{e^4}{16 \pi^2 q^4} \ y \
l^{\mu\nu} \ W_{\mu\nu} \, ,
\ea
where $y$ is the lepton fractional energy loss and $e$ denotes the
electron charge. The hadronic tensor depends on the hadron
structure, where there are important contributions from soft QCD
processes. For this reason the gauge/string theory duality becomes a
suitable tool for the calculation of this tensor, and therefore the
structure functions.

In this work we focus on the structure functions associated with
unpolarized vector mesons\footnote{The polarized structure functions
$b_{1,2,3,4}$ and $g_{1,2}$ \cite{Hoodbhoy:1988am} have been
obtained at strong coupling and in the planar limit in
\cite{Koile:2011aa,Koile:2013hba} from supergravity and in
\cite{Koile:2014vca} from superstring theory scattering amplitudes.
The latter gives the relevant behavior for small values of the
Bjorken parameter $x$.}. The corresponding hadronic tensor has the
form
\ba
W_{\mu\nu}&=& F_{1} (x,q^2) \, \eta_{\mu\nu} - \frac{F_{2}(x,q^2)}{P
\cdot q} \, P_{\mu}P_{\nu}  \, .
\ea
DIS is related to the forward Compton scattering (FCS) through the
optical theorem, which is an special case of the Cutkosky rules,
based on the fact that the $S$-matrix is unitary. It relates the
imaginary part of the FCS amplitude to the DIS amplitude. Then, the
tensor $T_{\mu\nu}$ is defined as
\begin{equation}\label{DIS50}
T_{\mu\nu}=i \, \langle P, {\cal
{Q}}|{\widehat{T}}(\tilde{J}_{\mu}(q) \, J_{\nu}(0))|P,  {\cal
{Q}}\rangle \, ,
\end{equation}
where $J_\mu$ and $J_\nu$ are the electromagnetic current operators,
${\cal {Q}}$ denotes the charge of the hadron, while ${\widehat{T}}$
represents time-ordered product. Tildes indicate the Fourier
transform. In terms of the optical theorem we can write
\begin{equation}\label{DIS51}
\textmd{Im}\,\tilde{F}_{j}=2\pi\,F_{j}\,,
\end{equation}
being $\tilde{F}_{j}$ the $j$-th structure function of the
$T_{\mu\nu}$ tensor, while $F_{j}$ is the one corresponding to the
$W_{\mu\nu}$ tensor.

At this point it is convenient to define the Bjorken parameter $x =
-q^2/(2 P \cdot q)$ for $q^2>0$, being its physical kinematic range
$0 \leq x \leq 1$. On the other hand, in the unphysical region for
$1 \ll x$ the product of the two electromagnetic currents in the
hadron can be written as an operator product expansion (OPE), in
terms of operators ${\cal {O}}_{n, k}$ multiplied by powers of
$(\Lambda^2/q^2)^{\gamma_{n,k}/2}$, where $n$ is the spin of ${\cal
{O}}_{n, k}$, while $\delta_{n,k}$, $\gamma_{n,k}$, and
$\Delta_{n,k}=\delta_{n,k} + \gamma_{n,k}$, represent the
engineering, the anomalous and the total scaling dimensions of the
operator, respectively \cite{Polchinski:2002jw}. Then, we can define
the twist of each operator as $\tau_{n,k} = \Delta_{n,k} - n$. The
relation with the physical parametric region $0 \leq x \leq 1$ is
given by a contour argument, which allows to connect the OPE with
the moments of the structure functions in the DIS process. Thus, the
$n$-moment of the $j$-th structure function can be expressed as the
sum of three contributions\footnote{We use the notation of
\cite{Polchinski:2002jw}, in particular equation (\ref{OPE}) is
similar to equation (27) of that reference but for mesons, i.e. the
second term has a factor $1/N$.}
\ba
M_n^j(q^2) &\approx& \frac{1}{4} \sum_k \, C_{n, k}^j \, A_{n, k} \,
\left(\frac{\Lambda^2}{q^2}\right)^{\frac{1}{2} \tau_{n, k}-1} +
\frac{1}{4} \sum_{{\cal {Q}}_p={\cal {Q}}} \, C_{n, p}^j \, A_{n, p}
\, \left(\frac{\Lambda^2}{q^2}\right)^{\tau_p-1} + \nn \\
&& \frac{1}{4} \, \frac{1}{N} \, \sum_{{\cal {Q}}_p \neq {\cal {Q}}}
\, C_{n, p}^j \, a_{n, p} \,
\left(\frac{\Lambda^2}{q^2}\right)^{\tau_p-1} \, , \label{OPE}
\ea
where the coefficients $C_{n, p}^j$ are dimensionless, while $A_{n,
p}$ and $a_{n, p}$ depend on the matrix elements of the operators
$<P, {\cal {Q}}|{\cal {O}}_{n, k}|P, {\cal {Q}}>$ for a hadron state
with four-momentum $P$ and charge ${\cal {Q}}$.

Let us briefly explain how different contributions behave in
equation (\ref{OPE}). We can study this equation for the photon
virtuality $q$ to be large, intermediate or small, in comparison
with the confinement scale $\Lambda$. At weak coupling the Feynman's
parton model gives a suitable description of hadrons, thus the
leading contribution comes from the first term. This contribution
only sums over terms associated with operators with the lowest twist
$\tau_{n,k} \approx 2$ at large $q^2$. In this case perturbative
methods of quantum field theory are suitable. On the other hand, the
second sum dominates at strong coupling and in the planar limit,
i.e. $1 \ll \lambda \ll N$. In this case protected double-trace
operators constructed from the protected single-trace ones have the
smaller twist at strong coupling. Therefore, the calculations can be
done by using the gauge/gravity duality, considering a forward
Compton scattering with the exchange of a single on-shell particle
between incoming and outgoing states. Within this regime exchange of
more than one intermediate state is suppressed by $1/N$ powers. The
third sum in equation (\ref{OPE}) becomes the leading one when $q^2
\geq \Lambda^2 N^{1/(\tau_{\cal {Q}}-\tau_c)}$. In this case
$\tau_{\cal {Q}}$ is the minimum twist of protected operators with
charge ${\cal {Q}}$, while $\tau_c$ is the minimum twist of all
electrically charged protected operators. The $1/N$ suppression of
the third sum is expected for mesons, while for glueballs there is a
$1/N^2$ suppression.

In addition, we should comment on the different parametric regions
in terms of $x$ and the 't Hooft coupling. For $1/\sqrt{\lambda} \ll
x \ll 1$ only supergravity states contribute since in this region
the ten-dimensional $s$-channel Mandelstam variable satisfies
${\tilde {s}} \ll 1/\alpha'$, where $\alpha'$ is the string
constant. When $\exp{(-\sqrt{\lambda})} \ll x \ll 1/\sqrt{\lambda}$
excited strings are produced and their dynamics becomes important.
The holographic dual calculation is derived from four-point
superstring theory scattering amplitudes. Finally, for the
exponentially small region the size of the excited string becomes
comparable with the AdS radius. In this case dual Pomeron techniques
are useful
\cite{Brower:2010wf,Brower:2006ea,Brower:2007xg,Brower:2007qh,Cornalba:2006xm,
Cornalba:2007zb,Watanabe:2012uc,Costa:2013uia,
Ballon-Bayona:2017vlm,Kovensky:2017oqs,Kovensky:2018xxa}. In
previous papers we have calculated $F_1$ and $F_2$ by considering
the FCS process with the exchange of a single intermediate state for
scalar and vector mesons
\cite{Koile:2011aa,Koile:2013hba,Koile:2014vca,Koile:2015qsa}. Then,
we have also calculated these functions by considering the exchange
of two intermediate states for glueballs \cite{Jorrin:2016rbx} and
for scalar mesons \cite{Kovensky:2016ryy} in the D3D7-brane system
of reference \cite{Kruczenski:2003be}. In both cases we found that
the large $N$ limit does not commute with the high energy one. By
considering the high energy limit first, which corresponds to the
physical situation, in the case of the pion we have obtained the
first moments of the structure function $F_2$ and compared them with
lattice QCD calculations
\cite{Best:1997qp,Brommel:2006zz,Chang:2014lva}, obtaining a
substantial improvement of the accuracy, namely: from 10.8$\%$ for a
single intermediate state \cite{Koile:2015qsa} to 1.27$\%$ in the
case of two intermediate states \cite{Kovensky:2016ryy}. Then, a
natural question is whether or not this effect also occurs in the
case of vector mesons. The present work answers it positively as we
shall explain in detail in the next sections.

For finite values of $N$ we can expand the structure functions of
mesons as follows
\begin{eqnarray}
F_j &=& f^{(0)}_j \,
\left(\frac{\Lambda^2}{q^2}\right)^{\tau_{in}-1} + \frac{1}{N} \,
f^{(1)}_j \, \left(\frac{\Lambda^2}{q^2}\right) + \frac{1}{N^2} \,
f^{(2)}_j \, \left(\frac{\Lambda^2}{q^2}\right) + \cdot \cdot \cdot
\label{Fj2}
\end{eqnarray}
where $\tau_{in}$ is the twist of the incident dual vector meson
state in type IIB supergravity, $f^{(n)}_j$'s indicate the structure
functions at order in $1/N^n$, with $j=1, 2$ and $n=1, \dots$.
Notice that in the definitions of $f^{(n)}_j$'s powers of
$\Lambda^2/q^2$ have been factored out. The index $n$ indicates the
number of exchanged intermediate on-shell states of the FCS Feynman
diagram. This corresponds to the number of hadrons in the final
state of the DIS process. From expression (\ref{Fj2}) one can easily
see that the high energy ($q^2 \gg \Lambda^2$) and the large $N$
limits do not commute. Moreover, by taking first the high energy
limit, since the power of $\Lambda^2/q^2$ in the first term is
larger than for the rest, it vanishes, and then in the $1/N$
expansion the second term dominates. We would like to emphasize that
$1/N$ corrections to the $F_1$ and $F_2$ structure functions for
scalar mesons were studied in \cite{Kovensky:2016ryy}, but not for
vector mesons. Therefore, we consider this calculation to be
important for the investigation of such limits beyond scalar
hadrons, since there are also lattice QCD results of the first
moments of $F_1$ for the rho meson to compare with
\cite{Best:1997qp}.

The work is organized as follows. In section 2 we discuss
generalities of DIS in the context of the D3D7-brane system at large
$N$. In section 3 we consider the $1/N$ expansion and obtain the
relevant Feynman-Witten diagram in the bulk theory. Also in this
section we develop the calculation of the $F_1$ and $F_2$ structure
functions for vector mesons. In section 4 we carry out the analysis
of our results. We focus on the calculation of the first moments of
the structure function $F_1$ of the rho meson and compare them with
the available results from lattice QCD.

%---------------------------------------------------------------------
%---------------------------------------------------------------------
\section{DIS in the D3D7-brane system}
%---------------------------------------------------------------------
%---------------------------------------------------------------------

DIS processes of charged leptons from scalar and vector mesons in
the D3D7-brane system have been studied in several papers
\cite{Koile:2011aa,Koile:2013hba,Koile:2014vca}, by considering the
large $N$ limit, which means that the final state has only a single
hadron. A more realistic calculation for vector mesons must include
$1/N$ corrections. This corresponds to  final multi-particle states.
In this work we consider $1/N$ corrections of DIS of charged leptons
from unpolarized vector mesons.

Firstly, we give a brief description of the D3D7-brane system. Let
us consider $N$ coincident D3-branes in type IIB superstring theory.
The corresponding near-horizon geometry is the AdS$_5\times S^5$
spacetime, with the metric
\bea
ds^2 = \frac{r^2}{R^2} \, \eta_{\mu\nu} dx^{\mu}dx^{\nu}+\frac{R^2}{r^2} d\Vec{Z}\cdot
d\Vec{Z} \, ,
\eea
where $\Vec{Z}$ are coordinates of the directions perpendicular to
the D3-branes, being the radial coordinate $r=|\Vec{Z}|$. The radius
of AdS$_5$ is $R = \left(4 \pi g_s N \al'^2\right)^{1/4}$,  where
$g_s$ is the string coupling. Now, one can add a D7-brane in the
probe approximation, at a distance $L = |\Vec{Z}|$ in the $(8,9)$
plane. The induced metric on the D7-brane is given by
\bea
ds^2 = \frac{\rho^2+L^2}{R^2} \, \eta_{\mu\nu} dx^{\mu}dx^{\nu} +
\frac{R^2}{\rho^2+L^2}d\rho^2+\frac{R^2\rho^2}{\rho^2 + L^2}
d\Omega_3^2 \, , \label{D3D7metric}
\eea
where $\rho^2 = r^2 - L^2$ and the angles contained in $\Omega_3$
span a three-sphere. For $L=0$ equation (\ref{D3D7metric}) gives the
AdS$_5\times S^3$ metric, otherwise the metric is only
asymptotically AdS$_5\times S^3$. This is the situation where the
conformal symmetry is preserved.

For $L>0$ the 3-7 quarks become massive, and meson type excitations
are energetically favored. Mesons correspond to excitations of open
strings ending on the D7-brane. The dynamics of these fluctuations
is described by the action
\bea
S_{D7} = -\mu_7 \int d^8\xi \sqrt{-\det\left(P\left[g\right]_{ab} +
2\pi \al' F_{ab} \right)} + \frac{\left(2 \pi \al'\right)^2}{2}\mu_7
\int P\left[C^{(4)}\right]\wedge F \wedge F \, , \label{D7action}
\eea
where $\mu_7 = [(2\pi)^7 g_s \al'^4]^{-1}$ is the D7-brane tension,
$\xi^a$ denotes the D7-brane coordinates, $g_{ab}$ stands for the
metric (\ref{D3D7metric}), and $P$ is the pullback of the background
fields on the D7-brane. The second term is the Wess-Zumino term.

It is possible to induce excitations in the transverse directions to
the D7-brane. These are two types of scalar excitations $\phi$ and
$\chi$, related to the $Z^5$ and $Z^6$ coordinates, respectively. On
the other hand, it is also possible to perturb the gauge fields
$F_{ab} = \der_a B_b - \der_b B_a$ on the Dirac-Born-Infeld (DBI)
action. In this case, there are three types of solutions for the
$B_a$ modes, related to the expansion the solutions in scalar or
vector spherical harmonics on $S^3$. The three classes of solutions
are \cite{Kruczenski:2003be}
\bea
\textrm{type I} &:& B_\m = 0 \, , \quad B_\rho = 0\, , \quad B_i =
\phi_I^{\pm}(\rho) \, \, e^{ik\cdot x} \,\, Y_i^{l\pm}(\Omega)\,, \label{BI}\\
\textrm{type II} &:&  B_\mu = \epsilon_\m \, \phi_{II}(\rho) \,
e^{ik\cdot x} \, Y^{l}(\Omega) \, , \,\,\, k\cdot \epsilon=0\,,
\quad B_\rho = 0 \, , \quad B_i=0 \, , \label{BII}\\
\textrm{type III} &:& B_\m =0 \, , \quad B_\rho = \phi_{III}(\rho) \
e^{ik\cdot x} \, Y^{l}(\Omega) \, , \quad B_i =
\Tilde{\phi}_{III}(\rho) \, e^{ik\cdot x} \, \nabla_i
Y^{l}(\Omega)\,. \label{BIII}
\eea
$Y^l(\Omega)$ and $Y^l_i(\Omega)$ are scalar and vector spherical
harmonics, respectively. Some of their properties are described in
the appendix and in references therein. Note that in this case type
I and III are scalar modes, while type II modes represent vector
fields from the (asymptotically) AdS perspective. The different
modes of the scalar and vector perturbations are shown in table 1,
together with their relevant quantum numbers.
{\renewcommand{\arraystretch}{1.4}
\begin{table}[h]
\begin{center}
\centering
\begin{tabular}{|c|c|c|c|c|}
\hline
Field & Type of field in $5D$ & Built from & $\Delta(l)$ & $SU(2)\times SU(2)$ irrep \\
\hline
$\phi,\chi$ & scalars & $\phi,\chi$ & $l+3$, $l\geq 0$ & $\left(\frac{l}{2},\frac{l}{2}\right)$ \\
\hline
$B_\mu$ & vector & $B_\mu^{II}$ & $l+3$, $l\geq 0$ & $\left(\frac{l}{2},\frac{l}{2}\right)$ \\
\hline
$\phi_I^{-}$ & scalar & $B_i^{I}$ & $l+1$, $l\geq 1$ & $\left(\frac{l+1}{2},\frac{l-1}{2}\right)$ \\
\hline
$\phi_I^+$ & scalar & $B_i^{I}$ & $l+5$, $l\geq 1$ & $\left(\frac{l-1}{2},\frac{l+1}{2}\right)$ \\
\hline
$\phi_{III}$ & scalar & $B_{i,z}^{III}$ & $l+3$, $l\geq 1$ & $\left(\frac{l}{2},\frac{l}{2}\right)$ \\
\hline
\end{tabular}
\caption{\small Some features of D7-brane fluctuations on the
AdS$_5\times S^3$ background relevant to this work. The integer $l$
indicates the $SO(4)\sim SU(2) \times SU(2)$ irreducible
representation (irrep) and it defines the corresponding Kaluza-Klein
mass. The relation between the scaling dimension of the associated
operator $\Delta$ and $l$ is also presented.}
\end{center}\label{Tabla1}
\end{table}

Beyond the quadratic order, the interaction Lagrangian for these
modes has been derived in reference \cite{Kovensky:2016ryy}.

Up to this point we have described the D3D7-brane system presented
in \cite{Kruczenski:2003be}, where the solutions were computed in
terms of hypergeometric functions. In the context of DIS from
mesons, one identifies the parameter that controls the separation
between the D7 and the D3-branes in the ($Z^5,Z^6$) plane with the
IR scale $\Lambda$ introduced as a cutoff in the radial direction to
induce confinement \cite{Polchinski:2002jw}. Thus, we take $L\sim
\Lambda R^2$. Therefore, the relevant interactions take place at
values of $\rho$ considerably larger than $L$, and in this region
the solutions are well approximated by the typical AdS$_5$
expressions in terms of Bessel functions, which we write in section
3. The AdS masses can only take discrete values. The presence of a
small but non-zero value of $L$ is important for the vertices we
will need to consider.

%---------------------------------------------------------------------
%
\subsection{One-particle exchange: the $N \rightarrow \infty$ limit}
%
%---------------------------------------------------------------------

For unpolarized vector mesons we shall study only the contributions
to the $F_i$ structure functions. These functions can be written in
terms of $W^{\mu\nu}$ and the vector $v^\m = \frac{1}{q}(P^\m +
\frac{q^\m}{2x})$ as
\bea
F_1(x,q^2) &=& \frac{1}{2} \left(g^{\m \n}-\frac{q^\m q^\n}{q^2}\right)
W_{\m \n} + 2 x^2v^\m v^\n  W_{\m \n} , \label{F1W}\\
F_2(x,q^2) &=& x\, \left(g^{\m \n}-\frac{q^\m q^\n}{q^2}\right)
W_{\m \n}
+ 12\, x^3 \, v^\m v^\n W_{\m \n} .\label{F2W}
\eea
%
%Note that the structure functions depend only on the virtual photon
%momentum transfer $q$ and the Bjorken parameter $x$.

The FCS amplitude can be derived by using the gauge/string theory
duality, by studying a four-point interaction with vector modes on
the D7-brane and gravi-photons related to current insertion on the
boundary as external states. This gauge field arises from a
particular decomposition of the graviton mode in ten dimensions:
\bea
\delta g_{mj} =  A_m \left(\rho,x\right) v_j\left(\Omega_3\right) \,
,
\eea
where $v_j$ are the Killing vectors on $S^3$, and $m =
(\mu, \rho)$.

The structure functions have been calculated in this context by
considering a single intermediate hadron state in
\cite{Koile:2011aa,Koile:2013hba}, obtaining the following results
at leading order
\bea
F_1(x,q^2) &=& A(x) \frac{1}{12x^3}(1-x)\,,
\\
F_2(x,q^2) &=& A(x) \frac{1}{6x^3}(1-x)\,,
\eea
with
\bea
A(x) = A_0 {\cal{Q}}^2\left(\frac{\mu^2_7(\al')^4}{\Lambda^8}\right)
\left(\frac{\Lambda^2}{q^2}\right)^{l+1} x^{l+6}(1-x)^{l}\,,
\eea
and $A_0=|c_i|^2|c_X|^2\,2^{6+2l}\left[\Gamma(3+l)\right]^2\pi^5$ is
a dimensionless constant. Also, $c_i$ and $c_X$ are the
normalization constants of the incident and intermediate dual hadron
states, respectively, while ${\cal{Q}}$ is the charge of the hadron,
associated to a $U(1)$ subgroup of the $S^3$ isometries.

These results are valid for DIS from mesons considered in the
context of the D3D7-brane system. However, it is important to keep
in mind that \cite{Koile:2011aa,Koile:2013hba} showed that
completely analogous formulas hold in the context of different
Dp-brane models, such as the D4D8$\overline{\mathrm{D8}}$-brane
system \cite{Sakai:2004cn} and the
D4D6$\overline{\mathrm{D6}}$-brane system \cite{Kruczenski:2003uq},
both in type IIA superstring theory. These models are very different
to each other, and each of them shares certain phenomenological
features with large-$N$ QCD. In consequence, it is reasonable to
expect that the qualitative form of the structure functions we just
described is universal, in the sense that it would hold in any
holographic Dp-brane model for mesons at strong coupling and in the
planar limit. Although in the present work we focus on the
D3D7-model, we expect this universality to hold also for the leading
one-loop correction, at least at the qualitative level.

%---------------------------------------------------------------------
%---------------------------------------------------------------------
\section{DIS in the $1/N$ expansion}
%---------------------------------------------------------------------
%---------------------------------------------------------------------

The non-planar $1/N$ corrections to $F_1$ and $F_2$ structure
functions for scalar mesons were studied in \cite{Kovensky:2016ryy},
and also for the ${\cal{N}}=4$ SYM theory glueball in reference
\cite{Jorrin:2016rbx}. From the point of view of DIS, they
correspond to processes with multi-particle final states. The first
non-trivial contribution comes from considering a two-hadron final
state in the hadronic tensor $W^{\m\n}_2$, which can be related to a
FCS process with two intermediate on-shell states, denoted as
$T^{\m\n}_2$. Writing this tensor in terms of the $U(1)$ conserved
current $J^{\m}$ we obtain \cite{Jorrin:2016rbx}
\bea
{\mathrm{Im}}\left(T^{\m\n}_2\right) &=& \pi \sum_{X_1,X_2}
\left<P,{\cal{Q}}\right|\Tilde{J}^\m (q) \left|X_1,X_2 \right>
\left<X_1,X_2\right|J^\n (0) \left|P,{\cal{Q}} \right>\nn \\
&=& \pi \sum_{M_2,M_3} \int \frac{d^3q'}{2E_{q'}(2\pi)^3}
\frac{d^3p'}{2E_{p'}(2\pi)^3} \left<P,{\cal{Q}}\right|\Tilde{J}^\m (q)
\left|X_1,X_2 \right>\left<X_1,X_2\right|J^\n (0) \left|P,{\cal{Q}}
\right>\nn \\
&=& 4\pi^3 \sum_{M_2,M_3} \int \frac{d^4q'}{(2\pi)^4}
\delta\left(M_2^2-q'^2\right)\delta\left(M_3^2-(P+q-q')^2\right)\nn\\
&& \qquad\qquad \times \left<P,{\cal{Q}}\right|J^\m (0) \left|X_1,X_2
\right>\left<X_1,X_2\right|J^\n (0) \left|P,{\cal{Q}} \right>\,,
    \label{Tmn}
\eea
where $X_1$ and $X_2$ are the intermediate states with momenta $p'$
and $q'$ respectively, as shown in figure 1. The current $J^\m$
matrix element is related to its Fourier transform as
\bea
\left<P,Q\right|\Tilde{J}^\m (q) \left|X_1,X_2 \right> =
\left(2\pi\right)^4
\delta^{(4)}\left(P+q-p'-q'\right)\left<P,Q\right|J^\mu (0)
\left|X_1,X_2 \right>\,.
\eea
Using the AdS/CFT duality, this current can be related to an
specific field in the bulk theory. In \cite{Jorrin:2016rbx},
considering $1/N$ corrections to DIS from a scalar meson we have
shown that the leading contribution to the DIS process with
two hadron final states is given by a specific Feynman diagram where
one of these two outgoing hadrons has the lowest twist. In the next
subsection we will explain the amplitude we need to calculate in the
case of a spin-1 meson.

%---------------------------------------------------------------------
%---------------------------------------------------------------------
\subsection{Leading diagram for vector mesons at order $1/N$}
%---------------------------------------------------------------------
%---------------------------------------------------------------------

We want to study the $1/N$ corrections to the DIS process from a
vector meson (for instance a rho meson), associated to a type II
vector mode $B^{II}_\mu$ on the D7-brane. Based on the results of
\cite{Kovensky:2016ryy}, the leading diagram is the $s$-channel one,
where the exchanged particle is the one with the lowest twist, $\tau
= \Delta - n$. This can be done by looking at table 1, which gives
the relevant quantum numbers of the different solutions.

Since the lowest $\tau$ is associated to the lowest $\Delta$, the
exchanged field should be the $\phi^{-}_{I}$ mode with $\tau =
\Delta = 2$. This is what has been done in \cite{Kovensky:2016ryy}.
However, the interaction term between $B^{II}_\mu$ and
$\phi^{-}_{I}$ modes given in \cite{Kovensky:2016ryy}
\beq
L_3 = -\mu_7(2\pi \al')^3 \sqrt{-g}
\frac{L}{\rho^2+L^2}\phi\left(F_{aJ}F^{aJ}-F_{a\m}F^{a\m} \right)\, \label{SIRgenerica},
\eeq
vanishes, due to the nature of the field solutions. It can be seen
from equations (\ref{BI}) and (\ref{BII}), that a type I scalar has
only angular components, while type II vectors have only $\mu$
components. Therefore, the interaction between these two modes
vanishes. The vector mode with $\tau=2$ does not contribute to the
DIS process either. This can be easily seen by analyzing the charge
of this vector associated with the 3-sphere. For $\tau=2$ we need to
consider $\Delta = 3$, but this implies that $l=0$, meaning that the
vector mode has no charge over the 3-sphere. Thus, there is no
interaction with the $A_m$ photon. Another possible interaction
could arise from the Wess-Zumino term in the low-energy action of
the D7-brane. This is because the gauge field is actually a
particular linear combination of the ten-dimensional graviton and RR
4-form perturbations. This is described in detail in
\cite{Kovensky:2017oqs,Kovensky:2018xxa}, where this type of vertex
has been used to study the antisymmetric contributions to the
hadronic tensor for glueballs and spin-$1/2$ hadrons. However, it
can be seen that the relevant angular integrals vanish in the
present case.

The next step is to consider the exchange of $\tau=3$ modes. There
are two possibilities:
\begin{enumerate}
\item $\phi,\chi$ scalars, with $\Delta=3$.
\item $B_\m$ vector, with $\Delta=4$.
\end{enumerate}
In the former case the perturbations have $l=0$, thus they are not
charged with respect to the $U(1)$ we are considering. Therefore,
the only possibility is the exchange of a type II vector mode with
$l=1$.

In the IR region, the relevant interaction includes two type II
modes (one associated to the $\tau=3$ mode we just discussed and the
other one with the incoming vector meson) and a scalar field, and
from \eqref{SIRgenerica} we see that it is described by
\cite{Kruczenski:2003be, Kovensky:2016ryy}
\beq
S_{\text{IR}} \approx \mu_7(2 \pi \al')^3 \frac{L}{R^4} \int d^8\xi \sqrt{g}\,
z^2 \phi\, F_{\m\n}F^{\m\n} \,. \label{S-IR}
\eeq
Note that we only keep the first term of the $L \ll \rho$ expansion,
where $\rho \simeq R^2/z$, being $z$ the Poincar\'e radial
coordinate of AdS$_5$.

~

For the UV vertex we have to consider the interaction between the
$A_m$ gauge field and two $B_\mu$ modes. The standard interaction is
\cite{Koile:2013hba}
\bea
S_{\text{UV}} = -\mu_7 (\pi\al')^2 i Q \int d^5x \sqrt{g}\,
A_\mu \left(B_\n^{*} F^{\m\n} - B_\n (F^{\m\n})^{*}\right)\,,
\eea
where we have already integrated over the $S^3$, being $Q$ the
charge of the vector mode, given by the eigenvalue equation $v^i
\der_i B_\m = i {\cal {Q}} B_\m$. Note that $Q$ does not have to
coincide with ${\cal{Q}}$, the charge of the original hadron. In
fact, one has to sum over all possible intermediate states, and in
particular all possible values of $Q$ (see appendix).

\begin{figure}[h!]
\centering
\includegraphics[scale=0.22]{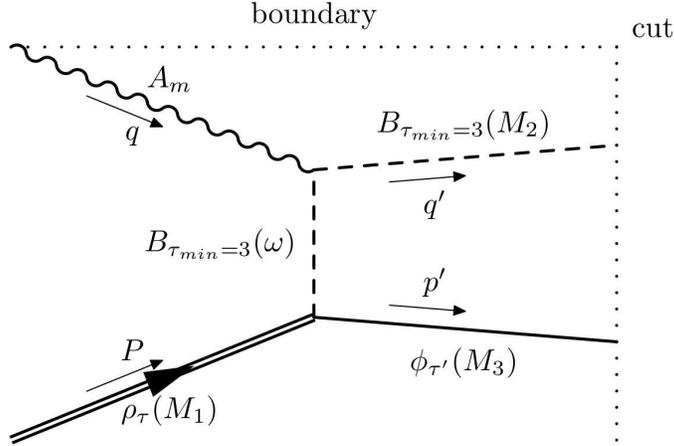}
\label{diagram} \caption{\small Feynman-Witten diagram associated to
the DIS process with two intermediate hadron states. The momentum
and the twist of each field are indicated, while the solutions are
given in section 2. The incident hadron is represented by a double
line, the intermediate spin-one modes are indicated by dashed lines,
while the solid line denotes a scalar mode. The vertical dotted line
represents the cut of the Cutkosky rule.}
\end{figure}

The diagram we need to calculate is shown in figure 1, where there
is an incoming photon with momentum $q_\m$, which interacts with a
massive $B_\m$ vector with momentum $q'_\m\,(q'^2=-M_2^2)$ and
$\tau_\textrm{min}=3$. This vector interacts with an incident rho
meson of momentum $P_\mu$, and with an outgoing scalar field with
momentum $p'_m\,(p'^2=-M_3^2)$ and conformal dimension $\tau'$.

In order to calculate the diagram we need the AdS$_5$ solutions of
the fields. In the axial gauge the gravi-photon solution is
\bea
A_\m(x,z) = n_\m \,e^{i q\cdot x} (qz) K_1(qz)\,,\quad A_z(x,z)=0\,,
    \label{Am}
\eea
while the fields on the probe D7-brane are given at small $L$ by the
following approximate expressions \cite{Koile:2013hba}
\bea
B_\m^*(x,z)&=& \frac{1}{\sqrt{N}} c^*_{B}\, \zeta_\m\, e^{-iq'\cdot
x} \sqrt{\frac{\Lambda}{ M_2}}\,z\, J_{2}(M_2z)\,,\nn \\
\rho_\m(x',z') &=& \frac{1}{\sqrt{N}} c_{\rho}\, \epsilon_\m\, e^{iP\cdot x'}
\sqrt{\frac{\Lambda}{M_1}}\,z J_{\tau-1}(M_1 z')\,,\nn \\
\phi^*(x',z') &=& \frac{1}{\sqrt{N}} c_{\phi}^*\,e^{-ip'\cdot x'}
\sqrt{\Lambda M_3}\,z J_{\tau'-1}(M_3 z')\,, \label{Fields}
\eea
where $c$'s are numerical constants, and we have also included the
polarization vectors. We have only written the AdS$_5$ solution, the
full ten-dimensional solution includes the 3-sphere contribution
that only have the product of the scalar spherical harmonics
$Y^l(\Omega_3)$. On the other hand, the propagator of the type II
vector mode is given by \cite{Raju:2011mp}
\bea
G_{\m\n}(x,x',z,z')&=& -\frac{i}{N}
\int \frac{d^4k}{(2\pi)^4}G^{(k)}(z,z') {\cal{T}}_{\m\n}\, e^{ik\cdot(x-x')}\nn\\
&=& -\frac{i}{N} \int
\frac{d^4k}{(2\pi)^4}\frac{d\omega^2}{2}\frac{(zz')J_{\tau-1}(\omega
z)J_{\tau-1}(\omega z')} {k^2+\omega^2-i\epsilon} {\cal{T}}_{\m\n}\,
e^{ik\cdot(x-x')}\,, \label{propagador}
\eea
where ${\cal{T}}_{\m\n} = \eta_{\m\n}+\frac{k_\m k_\n}{\omega^2 }$.
Recall that in our case of interest the exchanged mode has conformal
dimension $\tau_{min} = 3$.

Finally, we redefine the D7-brane fields as $\Phi \to
\frac{\Phi}{\sqrt{N}}$, such that they are canonically normalized in
terms of $N$. The $1/N$-power counting shows that the interaction
terms now scale as
\bea
S_{UV} \to \frac{1}{N} S_{UV} \, , \quad S_{IR} \to
\frac{1}{N^{3/2}} S_{IR} \, .
\eea
%

%---------------------------------------------------------------------
%---------------------------------------------------------------------
\subsection{Structure functions for vector mesons}
%---------------------------------------------------------------------
%---------------------------------------------------------------------

We now derive the FCS amplitude related to the one-point function
$n_\mu\left<X_1,X_2\right|J^\mu (0) \left|P,{\cal{Q}} \right>$.
Looking at the diagram of figure 1, the associated amplitude is
\bea
&& {\cal{A}} = - 8 i\,\mu_7^2 (\pi \al')^5 Q \frac{L}{R^4} (2\pi)^4
\delta^{(4)}(P+q-p'-q') \,
{\cal{I}}\, \int dz dz' \sqrt{g} \sqrt{g'} z'^2 \der'^{[\al}\rho^{\be]}(z') \phi^{*}(z') \times \nn \\
&&\left[B^{*\n}(z)\left(\der'_\al \der_\m G_{\n\be}(z,z')-\der'_\al
\der_\n G_{\m\be}(z,z') \right)-\der'_\al G_{\n\be}(z,z')
\left(\der_\m B^{*\n}(z)-\der^\n B^{*}_\m(z) \right)\right]A^\m(z)
\, ,  \nonumber \\
&& \label{Amplitud}
\eea
where $\der'_\al$ and $\der'_\be$ are derivatives with respect to
the primed coordinates. Also, we have already performed the
integrals in the variables $x$ and $x'$, which lead to the momentum
conservation relations
\bea
k_\m = q'_\m-q_\m = P_\m-p'_\m\,.
\eea
In equation (\ref{Amplitud}) ${\cal{I}}$ represents the integral of
the scalar spherical harmonics over the 3-sphere, which is given in
the appendix. Replacing the solutions (\ref{Am}), (\ref{Fields}) and
(\ref{propagador}) in the amplitude, and using the relations
(\ref{F1W}), (\ref{F2W}) and (\ref{Tmn}), the structure functions
$F_i$ ($i=1,2, L$) can be written as
\bea
F_i(q^2,x) &=& |\Tilde{c}|^2\, Q^2\, \mu_7^4\, L^2 \al'^{10}R^{8}\sum_{M_2,M_3}
\int \frac{d^3q'}{2E_{q'}(2\pi)^3} \frac{d^3p'}{2E_{p'}(2\pi)^3} \delta^{(4)}
\left( P+q-q'-p'\right)\nn\\
&& \times\Lambda^{3} \frac{ M_3}{M_1 M_2} q^{2} |C_t|^2 \,
F_i^T(x,P,q,q') \, , \label{FiIntegral} \eea
where $|\Tilde{c}|^2=|c_\phi|^2|c_B|^2|c_\rho|^2$, $C_t$ contains
the radial coordinate integrals, given by
\begin{align}
C_t = \int dz\, dz' \int d\omega \frac{\omega}{(q-q')^2+\omega^2}
\times \quad \quad \quad \quad & \nn \\
\left[ z'^4 J_{\tau-1}(M_1z') J_{\tau'-1}(M_3z') J_2(\omega
z')\right]& \left[ z^2 K_1(qz) J_{2}(M_2z) J_2(\omega z) \right] \,.
\label{Ct1}
\end{align}
In order to obtain $F_i^T$, the factor in \eqref{FiIntegral} which
depends only of the tensor contractions, we need to calculate
$J^T_\m J_\n^{T*}$, with
\bea
J_\m^T = \left[\zeta^{*\n} \left(k_\m {\cal{T}}_{\n \be}-k_\n
{\cal{T}}_{\m \be}\right) + {\cal{T}}_{\n \be} \left(q'_\m
\zeta^{*\n}-q'^\n \zeta^{*}_\m\right)\right]  k_\al
P^{[\al}\epsilon^{\be]}\,.
\eea
Recall that $B^{*}_\mu$ is an intermediate state, thus we need to
sum over the outgoing vector polarizations $\zeta_\m$
\bea
\sum_{\lambda} \zeta_\m \zeta_\n^{*} =  -q'^2\eta_{\m\n}+q'_\m q'_\n
\,. \label{sumzeta}
\eea
Since we are only interested in the unpolarized structure functions,
we also average over the polarization vector of the incoming hadron
\bea
\overline{\epsilon_\m \epsilon_\n^{*}} =
\frac{1}{3}\left(-P^2\eta_{\m\n}+P_\m P_\n \right) \,.
\label{sumeps}
\eea
By comparing $J^T_\m J_\n^{T*}$ with equations (\ref{F1W}) and
(\ref{F2W}) we obtain the following expressions
\bea
F_1^T &=& \frac{1}{24 q^2 x^2}P^2 \left[128 x^4 \left(q'\right)^2
\left(P\cdot q'\right)^4+128 x^3 \left(q'\right)^2 q\cdot q'
\left(P\cdot q'\right)^3+8 q^6 x \left(q'\right)^2 P\cdot q'
\right.\nn \\
&&\left.  +40 q^4x^2 \left(q'\right)^2 \left(P\cdot q'\right)^2 -8
q^4 x^2 \left(q\cdot q'\right) \left(P\cdot q'\right)^2+32 q^4 x
\left(q'\right)^4 P\cdot q'\right.\nn \\
&&\left.-16 q^4 x \left(q\cdot q'\right)^2 P\cdot
q' +28 q^4 x \left(q'\right)^2 \left(q\cdot q'\right) \left(P\cdot q'\right) -16 q^2 x^4
\left(P\cdot q'\right)^4\right.\nn \\
&&\left.
+128 q^2 x^3 \left(q'\right)^2 \left(P\cdot
q'\right)^3-48 q^2 x^3 \left(q\cdot q'\right) \left(P\cdot q'\right)^3+32 q^2 x^2
\left(q'\right)^4 \left(P\cdot
q'\right)^2 \right.\nn \\
&&\left. -40 q^2 x^2 \left(q\cdot q'\right)^2 \left(P\cdot
q'\right)^2+128 q^2 x^2 \left(q'\right)^2 \left(q\cdot q'\right) \left(P\cdot
q'\right)^2+2 q^8 \left(q'\right)^2
\right.\nn \\
&&\left.+9 q^6 \left(q'\right)^4
-2 q^6 \left(q\cdot q'\right)^2-2 q^6 \left(q'\right)^2
q\cdot q' \right] \\
&&\nonumber  \\
F_2^T &=&\frac{1}{12 q^2 x}P^2\left[384 x^4 \left(q'\right)^2
\left(P\cdot q'\right)^4+384 x^3 \left(q'\right)^2 \left(q\cdot q'\right)
\left(P\cdot q'\right)^3\right.\nn \\
&&\left.
+64 x^2 \left(q'\right)^2 \left(q\cdot
q'\right)^2 \left(P\cdot q'\right)^2 +8 q^6 x \left(q'\right)^2 P\cdot q'+96 q^4 x^2
\left(q'\right)^2 \left(P\cdot q'\right)^2\right.\nn \\
&&\left.
-8 q^4 x^2 \left(q \cdot q' \right)
\left(P\cdot q'\right)^2+32 q^4 x \left(q'\right)^4 \left(P\cdot q'\right)
-16 q^4 x \left(q\cdot q'\right)^2 \left(P\cdot q'\right) \right.\nn \\
&&\left.+76 q^4 x
\left(q'\right)^2 \left(q\cdot q'\right) \left(P\cdot q'\right)-48 q^2 x^4 \left(P\cdot
q'\right)^4+384 q^2 x^3 \left(q'\right)^2 \left(P\cdot
q'\right)^3\right.\nn \\
&& \left. -144 q^2 x^3 \left(q\cdot q'\right) \left(P\cdot q'\right)^3 + 32 q^2
x^2 \left(q'\right)^4 \left(P\cdot q'\right)^2 -136 q^2 x^2
\left(q\cdot q'\right)^2 \left(P\cdot q'\right)^2 \right.\nn
\\
&&\left. +384 q^2 x^2 \left(q'\right)^2 \left(q\cdot q'\right) \left(P\cdot
q'\right)^2-32 q^2 x \left(q\cdot q'\right)^3 \left(P\cdot q'\right) +64 q^2 x
\left(q'\right)^2 \left(q\cdot q'\right)^2 \left(P\cdot q'\right)\right.\nn \\
&&\left.+2 q^8
\left(q'\right)^2+9 q^6 \left(q'\right)^4 -2 q^6 \left(q\cdot q'\right)^2-2
q^6 \left(q'\right)^2 \left(q\cdot q'\right)+8 q^4 \left(q'\right)^2 \left(q\cdot
q'\right)^2 \right]  \,.
\eea
In order to calculate the integrals in (\ref{Ct1}) we need to use a
few reasonable approximations, in a similar way as in
\cite{Kovensky:2016ryy,Jorrin:2016rbx}. The main assumption is that
$\Lambda$ and the masses of the hadrons are small in comparison with
the momentum of the virtual photon. The IR integral selects the mass
of the exchanged field as follows \cite{Jorrin:2016rbx}
\begin{align}
\int_0^{\Lambda^{-1}} dz'z'^4J_{\tau-1}(M_1z')J_{\tau'-1}(M_3 z')J_2(\omega z')\approx &\\
\frac{1}{\Lambda^3}\frac{1}{\sqrt{M_1M_3}}[(-1)^\al
&\delta(\omega-(M_1+M_3))+(-1)^\be \delta(\omega-(M_1-M_3))]\,, \nn
\end{align}
for some integers $\alpha$ and $\beta$. The integral leads to
$\omega=|M_1\pm M_3|$. Then, the UV integral can be obtained by
expanding $J_2(\omega z)\approx \omega^2 z^2/8$ for $\omega \ll q$,
and taking the upper limit as infinity since $K_1$ decays quickly in
the bulk. We obtain
\bea
\int_0^{\infty} dz\,\frac{\omega^2}{8}z^4\,K_{1}(qz)J_{2}(M_2 z)
\approx \frac{6M_2^2 q\, \omega^2}{(M_2^2+q^2)^4}\,.
\eea
With these two equations we can obtain $C_t$, after noticing that
the leading term comes from $\omega=M_1-M_3$
\cite{Kovensky:2016ryy}, we obtain
\bea
|C_t|^2 = \frac{q^236M_2^4}{(M_2^2+q^2)^8} \frac{1}{\Lambda^6}
\frac{(M_1-M_3)^6}{M_1M_3}
\frac{1}{\left((q-q')^2+(M_1-M_3)^2\right)^2}\,.
\eea
The next step is to integrate over the on-shell momenta $\Vec{p'}$
and $\Vec{q'}$, and sum over the corresponding masses\footnote{For
more details we refer the reader to reference
\cite{Kovensky:2016ryy}.}. The final results for the structure
functions are
\bea
F_1(x,q^2) &=&\frac{1}{\lambda N}  C \left(\frac{M_1}{\Lambda}
\right)^6 \frac{\Lambda^2}{q^2}\,\frac{1}{2}\, x^3(1-x)^3
 \\
F_2(x,q^2) &=& \frac{1}{\lambda N}  C \left(\frac{M_1}{\Lambda}
\right)^6 \frac{\Lambda^2}{q^2}\, x^3(1-x)^3(4-3x)
\\
F_L(x,q^2) &=& \frac{1}{\lambda N}  C \left(\frac{M_1}{\Lambda}
\right)^6 \frac{\Lambda^2}{q^2}\,4\, x^3(1-x)^4\,,
\eea
where $C$ is a numerical constant.

We expect qualitatively similar results to hold in the context of
different Dp-brane models.

%---------------------------------------------------------------------
%---------------------------------------------------------------------
\section{Discussion and conclusions}
%---------------------------------------------------------------------
%---------------------------------------------------------------------

We have obtained the $1/N$ corrections to the $F_1$, $F_2$ and $F_L$
structure functions corresponding to vector mesons, using the
gauge/gravity duality. Motivated by previous work for ${\cal {N}}=4$
SYM theory glueballs, and particularly for scalar mesons in the
D3D7-brane system, the idea is to investigate how two very different
limits behave, namely: the large $N$ limit in comparison with the
high energy limit ($\Lambda^2/q^2 \rightarrow 0$). Our first result
is that they do not commute for the vector mesons. Then, since the
physical way to consider these limits implies to take first the high
energy one, followed by the large $N$ limit, we find that in this
situation the third term in the expansion of equation (\ref{OPE})
dominates the moments of the structure functions. This is a very
interesting result which says that at strong coupling this third
term becomes the leading one for $q^2 \geq \Lambda^2
N^{1/(\tau_{\cal {Q}}-\tau_c)}$, where $\tau_{\cal {Q}}$ and
$\tau_c$ are the minimum twist of protected operators with charge
${\cal {Q}}$ and the minimum twist of all electrically charged
protected operators, respectively.

This is similar to what happens for the glueball in the IR deformed
version of ${\cal{N}}=4$ SYM
\cite{Polchinski:2002jw,Jorrin:2016rbx}, where the process is given
in terms of closed string modes and at strong coupling one finds
that the $1/N$ result dominates for $q^2 \geq \Lambda^2
N^{2/(\tau_{\cal {Q}}-\tau_c)}$. However, note that in the present
case, i.e. for mesons, the correction is proportional to the $1/N$
instead of $1/N^2$, as expected. Thus, at large $N$ the critical
value for the photon virtuality $q$ where this happens is much
smaller. The same occurs for the results presented in
\cite{Kovensky:2016ryy} for the case of scalar mesons.

The physical implication of this result is that, at strong coupling,
for the above energy range DIS is dominated by a two-hadron final
state. From the viewpoint of the FCS process it corresponds, through
the optical theorem, to a situation where there are two intermediate
hadron states. The structure functions $F_1$ and $F_2$ behave as
$(1-x)^3$ as $x$ approaches 1. We have also considered the
longitudinal structure function $F_L$ which behaves as $(1-x)^4$ as
$x$ approaches 1. We should notice that, although the states as well
as the interactions for the vector and scalar mesons are different,
all these structure functions have the same dependence on
$\Lambda^2/q^2$, $1/N$, $1/\lambda$ and $M_1/\Lambda$ as in the case
of scalar mesons in the D3D7-brane system.

It is important to consider the moments of the structure functions
defined as
\begin{equation}
M_n [F_i] = \int_0^1 dx \, x^{n-1} F_i(x,q^2) \ ,
\end{equation}
where $F_i$ can be $F_1$ and $F_2$ in this case.

Several moments of these structure functions have been calculated in
reference \cite{Koile:2015qsa} in the large $N$ limit, i.e. by
considering a single intermediate hadron state in the FCS process.
This has been done for the first moments of the structure function
$F_2$ in the case of the pion as well as for $F_1$ of the rho meson.
In \cite{Koile:2015qsa} we have compared these results with lattice
QCD data from references
\cite{Best:1997qp,Brommel:2006zz,Chang:2014lva} for the pion,
associated with the lightest pseudoscalar mode. In addition, in the
case of the rho meson associated to the $l=2$ spin-$1$ mode of the
type II solutions, the comparison has been made with respect to
results from lattice QCD of \cite{Best:1997qp}. The best fit for the
case of the pion leads to results with an accuracy of $10.8 \%$ or
better, while in the case of the rho meson the accuracy is of $18.5
\%$ or better \footnote{There is a mistake in table 4 of reference
\cite{Koile:2015qsa} that we have corrected here. The original
errors presented in that reference were overestimated.}, for the
D3D7-brane system. In \cite{Koile:2015qsa} also the Sakai-Sugimoto
model of the D4D8$\mathrm{\overline{D8}}$-brane system and the
D4D6$\mathrm{\overline{D6}}$-brane system, both in type IIA string
theory, have been considered for FCS with a single intermediate
exchanged state. The next step has been done in
\cite{Kovensky:2016ryy} where we have considered the leading $1/N$
corrections to the structure functions. The accuracy is notoriously
enhanced to $1.27 \%$ for the scalar mesons in the D3D7-brane system
in this case. It leads to a natural question which is whether for
vector mesons the accuracy of the fit can also be substantially
improved by considering $1/N$ corrections.

In order to investigate this point we have carried out the best fit
of the structure $F_1$ including $1/N$ corrections in comparison
with lattice QCD data from \cite{Best:1997qp}. Recall that the
results of the present work have been obtained in the type IIB
supergravity regime, i.e. where $1/\sqrt{\lambda} \ll x < 1$, which
means that for the calculation of the moments we have integrated our
result for the functions between $x=0.1$ and $x=1$. On the other
hand, we also need to carry out the integration over the range
$\exp{(-\sqrt{\lambda})} \ll x \ll 1/\sqrt{\lambda}$, where we
assume that the behavior of the structure functions is similar to
the behavior shown in \cite{Koile:2014vca} and used in
\cite{Koile:2015qsa}, i.e. $F_L^{small-x} \propto x^{-1}$. We
support this assumption on the fact that, in the $1/\sqrt{\lambda}
\ll x < 1$ range the difference between the large-$N$ calculation
(where there is only a single on-shell hadron state exchanged in the
FCS process) and the $1/N$ calculation (where the leading
Feynman-Witten diagram has two on-shell hadron states exchanged) is
that in the former the dependence with the photon virtuality and the
Bjorken parameter is given by $\tau_{in}$ corresponding to the
incident meson, while in the later the $q^2$ and $x$ dependence is
determined by $\tau_{min}$. This corresponds to one of the lowest
conformal dimension from the supergravity excitations. However, for
$\exp{(-\sqrt{\lambda})} \ll x \ll 1/\sqrt{\lambda}$ things are
different, namely: the calculation from the two-open and two-closed
strings scattering amplitudes is independent of $\tau_{in}$. Thus,
we assume that in this low-$x$ regime the genus-zero result from
type IIB superstring theory should not be very different with
respect to a much more complicated calculation on the torus. Then,
for this string theory regime of the holographic dual calculation we
use the expressions for the structure functions at tree level from
\cite{Koile:2015qsa}. For our numerical calculation at low-$x$ we
consider that the integration for the moments is performed between
$x=0.0001$ and $x=0.1$ as before
\cite{Koile:2015qsa,Kovensky:2016ryy}. Then, we split each structure
function in two parts, each of one having a dimensionless constant
to be fixed by fitting with respect to lattice QCD data
\cite{Best:1997qp}. There is a constant $C_1$ multiplying the
low-$x$ $F_1$ function. In addition, there is a second constant
$C_2$ on the large-$x$ $F_1$ function.

Results of the first moments of $F_1$ of the rho meson are presented
in table 2.
%%%%
\begin{table}[h]
\def\arraystretch{1.5}
\begin{center}
\begin{tabular}{|c|c|c|c|}
\hline
Model / Moment & $M_2(F_1)$ & $M_3(F_1)$ & $M_4(F_1)$ \\
\hline
Lattice QCD & 0.1743 & 0.074 & 0.035 \\
\hline
D3D7 ($N\rightarrow \infty$) & 0.1753 & 0.060 & 0.039 \\
\hline
Percentage error & -0.6 & 18.5 & -12.8 \\
\hline
D3D7 ($1/N$) & 0.1750 & 0.065 & 0.038 \\
\hline
Percentage error & -0.39 & 12.5 & -9.6 \\
\hline
\end{tabular}
\caption{\small Comparison of our results for the first moments of
the structure function $F_1$ of the lightest vector meson for a
suitable choice (best fitting) of the normalization constants with
respect to the results of the lattice QCD simulations in
\cite{Best:1997qp} and in comparison with previous results presented
in \cite{Koile:2015qsa}. Uncertainties in the lattice QCD
computations are omitted.}\label{momentstablepi}
\end{center}
\end{table}
%%%
The values we obtain for the constants are $C_1=0.0087$ and
$C_2=32.1939$. Note that they are of the same order as the ones
found in our previous work \cite{Koile:2015qsa} in the large $N$
limit, for which the constants associated with the small-$x$ $F_1$
and with the large-$x$ $F_1$ of the rho meson are $0.012$ and
$78.07$, respectively.

Figure 2 shows the structure function $F_1$ as a function of $x$.
The blue bell-shaped curve indicates the $1/N$ calculation of this
work. The black dashed bell-shaped line corresponds to the case
obtained in reference \cite{Koile:2015qsa} for the large $N$ limit.
For small-$x$ values we use the result from \cite{Koile:2014vca},
leading to the monotonically decreasing curves. The difference
between the two curves at small values of the Bjorken parameter
comes from the slightly different constants which correspond to the
best fit developed in each situation.
\begin{figure}[h!]
\centering
\includegraphics[scale=1.,trim = 0 0 0 0 mm]{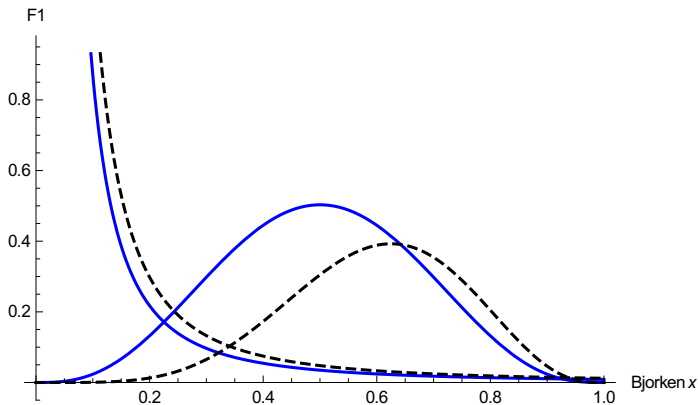}
\caption{\small $F_1$ as a function of $x$. We display the leading
results at low and moderate ($0.1 < x \leq 1$) values of $x$. Dashed
curves represent the $F_1$ computed for $N\to\infty$, while the blue
ones correspond to the $1/N$ corrections. The constants $C_1$ and
$C_2$ are those which give the moments of $F_1$ shown in table 2.}
\label{First-Moments-Pion}
\end{figure}
There is an important improvement with the inclusion of the leading
$1/N$ correction. As it happened in the scalar case, the location of
the maximum is shifted to the left with respect to the results
obtained in the planar limit, matching better the phenomenological
expectations\footnote{See for example references
\cite{Mankiewicz:1988dk,Sun:2017gtz}.}.

From table 2 we can appreciate the enhancement of the accuracy of
the moments of $F_1$ for the case of the rho meson which goes from
$18.5 \%$ for one particle exchange in the FCS, down to $12.5 \%$ in
the leading $1/N$ contribution, i.e. for the exchange of two
intermediate on-shell states. This is very important because it
confirms the trend found previously for glueballs in ${\cal {N}}=4$
SYM theory \cite{Jorrin:2016rbx} and for the scalar mesons of the
${\cal {N}}=2$ SYM theory from the D3D7-brane system
\cite{Kovensky:2016ryy}. Thus, it indicates that in order to infer
realistic conclusions for physical systems it is crucial to consider
the $1/N$ expansion of the observables, and consider first the large
momentum transfer limit. Possibly, this behavior can be identified
in other physical process, and although we have restricted our
investigation to strongly coupled gauge theories it would be very
interesting to investigate the relations between the large $N$ limit
and the high energy one for gauge field theories at perturbative
level.

~

~

%%%%%%%%%%%%%%%%%%%%%%%%%%%%%%%%%%%%%%%%%%
%
\centerline{\large{\bf Acknowledgments}}
%
%%%%%%%%%%%%%%%%%%%%%%%%%%%%%%%%%%%%%%%%%%

~

N.K. acknowledges kind hospitality at the Institut de Physique
Th\'eorique, CEA Saclay, and at the Institute for Theoretical
Physics, University of Amsterdam, during the completion of this
work. G.M. acknowledges kind hospitality at the International Center
for Theoretical Physics, Trieste, where part of this work has been
done. This work has been supported by the National Scientific
Research Council of Argentina (CONICET), the National Agency for the
Promotion of Science and Technology of Argentina (ANPCyT-FONCyT)
Grants PICT-2015-1525 and PICT-2017-1647, and the CONICET Grant
PIP-UE B\'usqueda de nueva f\'{\i}sica.

\newpage

%---------------------------------------------------------------------
%---------------------------------------------------------------------

\section*{Appendix: Angular integrals}
\addcontentsline{toc}{section}{Appendix: Angular integrals}

Scalar spherical harmonics on the 3-sphere belong to the $(l/2,l/2)$
representation of $SU(2)\times SU(2) \equiv SO(4)$, where $l$ is a
non-negative integer, while $-l/2\leq m\,,n \leq l/2$. They form a
basis of eigenfunctions of the Laplace operator on the sphere,
\bea
\nabla^2 Y^{m,n}_l = -l (l +2)Y^{m,n}_l\,,
\eea
and satisfy the orthogonality relation
\bea
\int_{S^3} (Y^{m,n}_l)^* \, Y^{m',n'}_{l'} =
\delta_{ll'}\delta_{mm'}\delta_{nn'}\,.
\eea
where
\bea
\left(Y^{m,n}_l\right)^* = \left(-1\right)^{m+n}Y^{-m,-n}_l\,.
\eea
The integral we want to calculate has three of scalar spherical
harmonics. An analytic expression for this type of integrals can be
found in \cite{Cutkosky:1983jd,Aharony:2006rf}, and it reads
\bea
\int_{S^3} Y^{m,n}_l Y^{m',n'}_{l'} Y^{m'',n''}_{l''} =
R_1(l,l',l'')
\begin{pmatrix}
\frac{l}{2} & \frac{l'}{2} & \frac{l''}{2} \\
m & m' & m''
\end{pmatrix}
\begin{pmatrix}
\frac{l}{2} & \frac{l'}{2} & \frac{l''}{2} \\
n & n' & n''
\end{pmatrix}\,,
\eea
where we have included the 3j-symbols, and the function $R_1$ is
defined as
\bea
R_1(x,y,z)=\frac{(-1)^\sigma}{\pi}
\sqrt{\frac{(x+1)(y+1)(z+1)}{2}}\,,\quad \textrm{with}\,\sigma
=\frac{(x+y+z)}{2}
\eea
and must satisfy the triangle inequality $|x-z|\leq y \leq x+z$ in
order to be non-vanishing. Also, the 3j-symbols must satisfy the
physical condition: $m+m'+m''=n+n'+n''=0$.

In our case of interest, i.e. the UV vertex of the diagram shown in
figure 1, the incoming state is a rho meson with quantum numbers
$(l,m,n)$, the exchanged particle is a vector mode with
$(l',m',n')$, and the outgoing state is the scalar $\phi$ with
$(l'',m'',n'')$. The leading diagram is the one where $\Delta'= 4$,
fixing $l' = 1$ and $-1/2\leq m'\,,n' \leq 1/2$. The rho meson is a
hadron which can have positive, null or negative charge. In the
D3D7-brane system, the charge is given by the $R$-symmetry
associated to the $R$ sector of the $SU(2)_L\times SU(2)_R$. This
implies that $l=2$, while $m$ can take three values $1,0,-1$ for a
positive, null or negative charged meson respectively. Therefore, we
have the following product of 3j-symbols
\bea
R_1(2,1,l'')
  \begin{pmatrix}
    1 & \frac{1}{2} & \frac{l''}{2} \\
    m & \pm \frac{1}{2} & -m\mp\frac{1}{2}
  \end{pmatrix}
  \begin{pmatrix}
    1 & \frac{1}{2} & \frac{l''}{2} \\
    n & \pm \frac{1}{2} & -n\mp \frac{1}{2}
  \end{pmatrix}\,,
  \label{3jsymbol2}
\eea
where $m$ can be $1,0,-1$. For any of these values, equation
(\ref{3jsymbol2}) does no vanish only if $l''=1,2,3$. We want to sum
over all possible exchanged and outgoing states, this is a sum over
$m',n'=\pm 1/2$, and over $l''$. The final result is the same for
the three values of $m$, and it reads
\bea
\sum_{m',n'=-1/2}^{1/2}\,\sum_{l''=1}^3\left[R_1(2,1,l'')
  \begin{pmatrix}
    1 & \frac{1}{2} & \frac{l''}{2} \\
    m & m' & -m-m'
  \end{pmatrix}
  \begin{pmatrix}
    1 & \frac{1}{2} & \frac{l''}{2} \\
    n & n' & -n-n'
  \end{pmatrix}\right]^2 = \frac{6}{ \pi^2}\,,
  \label{3jsymbol3}
\eea
which is the value of ${\cal{I}}$ present in equation
(\ref{Amplitud}).

%---------------------------------------------------------------------
%---------------------------------------------------------------------

\newpage

%%%%%%%%%%%%%%%%%%%%%%%%%
% BIBLIOGRAPHY
%%%%%%%%%%%%%%%%%%%%%%%%%

\end{document}